\documentclass[pdflatex,bst/sn-nature]{sn-jnl}

\usepackage{graphicx}%
\usepackage{multirow}%
\usepackage{amsmath,amssymb,amsfonts}%
\usepackage{amsthm}%
\usepackage{mathrsfs}%
\usepackage[title]{appendix}%
\usepackage{xcolor}%
\usepackage{textcomp}%
\usepackage{manyfoot}%
\usepackage{booktabs}%
\usepackage{algorithm}%
\usepackage{algorithmicx}%
\usepackage{algpseudocode}%
\usepackage{listings}%
\usepackage{hyperref}%
\usepackage{lineno}%
\usepackage{setspace}%
\usepackage{float}%

\begin{document}

\title[Article Title]{Polarization-Insensitive Integration of Nanoparticle-on-a-Slit Cavities with Dielectric Waveguides for On-chip Surface Enhanced Raman Spectroscopy}


\author[1]{\fnm{Javier} \sur{Redolat}}\email{jaredque@ntc.upv.es}

\author[1]{\fnm{Daniel} \sur{Arenas-Ortega}}\email{arenas2222@hotmail.com}

\author[2]{\fnm{Ángela} \sur{Barreda}}\email{abarreda@ing.uc3m.es}

\author[1]{\fnm{Amadeu} \sur{Griol}}\email{agriol@ntc.upv.es}

\author*[1]{\fnm{Elena} \sur{Pinilla-Cienfuegos}}\email{epinilla@ntc.upv.es}

\author*[1]{\fnm{Alejandro} \sur{Martínez}}\email{amartinez@ntc.upv.es}

\affil[1]{\orgdiv{Nanophotonics Technology Center}, \orgname{Universitat Politècnica de València}, \orgaddress{\street{ Camino de Vera, s/n, Edificio 8F}, \city{Valencia}, \postcode{46022},\country{Spain}}}

\affil[2]{\orgdiv{Department Electronic Technology}, \orgname{Carlos III University of Madrid}, \orgaddress{\street{ Avda. de la Universidad 30}, \city{Leganés}, \postcode{ 28911},\country{Spain}}}

\abstract{Amongst the available plasmonic nanostructures, nanoparticle-on-a-mirror (NPoM) cavities - consisting of metal nanoparticles separated from a metal mirror by a molecular-size monolayer - provide the ultimate light confinement in gaps even below 1 nm. A variation of the NPoM cavity is the nanoparticle-on-a-slit (NPoS) configuration, where the nanoparticle is placed on a functionalized narrow slit created on a metal plate so that there are two nanometric-scale gaps for plasmonic localization. Interestingly, the NPoS cavity can also perform as a dual dipole antenna to localize both infrared and visible light, which is useful in molecular optomechanics \cite{roelli:2024}. For many applications, it is desirable to integrate such cavities on a chip and provide access to (and collection from) the hot spots via photonic integrated waveguides. In this work, we propose, design, and experimentally demonstrate the efficient integration of NPoS plasmonic cavities with dielectric waveguides on a silicon-based chip. To this end, we use silicon-nitride slot waveguides and show that both the fundamental TE and TM modes can be used to drive the cavity, making our device polarization-independent. We demonstrate our concept by performing surface-enhanced Raman spectroscopy of BPT molecules on a chip fabricated by standard silicon fabrication tools mixed with the deterministic positioning of gold nanospheres on the gap of a plasmonic dipole antenna. }



\keywords{Surface-Enhanced Raman Spectroscopy (SERS), hybrid plasmonic-photonic circuits, plasmonic antennas, silicon photonics, self-assembled monolayers, nano-particle on a mirror cavity, integrated dielectric waveguides, on-chip spectroscopy, molecular optomechanics.}

\maketitle


\section*{Introduction}\label{sec1}
Recent advances in the field of nanophotonics have enabled the development of miniaturized optical devices for application in a wide range of fields such as biology \cite{Punj:2013}, quantum optics \cite{Javadi:2015}, photovoltaics \cite{Atwater:2010}, and sensing \cite{Herkert:2021}. A common feature of all such devices is the localization of light in regions much smaller than the wavelength, which maximizes light-matter interaction and its resulting nonlinear effects \cite{Koenderink:2016}. Subwavelength localization usually requires using metals that, when appropriately nanostructured, can support plasmonic resonances with extreme field localization beyond the diffraction limit in nanometric-size hot spots \cite{Schuller:2010, Brongersma:2015}. A well-known application where plasmonic structures provide orders-of-magnitude improvement in performance is Raman spectroscopy. Here, the use of nanostructured metals can lead to the maximization of the Raman effect, ultimately leading to the detection of single molecules in a field that is commonly referred to as surface-enhanced Raman spectroscopy (SERS) \cite{nie:1997,Kneipp:1997,Langer:2019,maccferri:2021}. 

In principle, plasmonics ultimately enables light localization in regions as tiny as a single atom or molecule \cite{Emboras:2015}. To this end, one of the most adequate plasmonic nanostructure in terms of fabrication viability is the so-called nanoparticle-on-a-mirror (NPoM) cavity \cite{Ciraci:2012,baumberg:2019}. In an NPoM cavity, a metal nanoparticle (NP) is placed on a metallic surface previously functionalized by a self-assembled monolayer (SAM). As a result, light can be tightly confined in the nanometric-scale gap between the NP and the surface upon proper illumination, overcoming the diffraction limit. Besides being highly interesting for SERS, the extreme features of NPoM cavities lead to the realization of ultrathin absorbers \cite{Moreau:2012} or the observation of intriguing phenomena such as the formation of picocavities \cite{benz:2016}. In a standard NPoM cavity, NPs are usually deposited randomly by drop casting and the underlying metal layer acts merely as an optical mirror \cite{baumberg:2019}. However, we may think of structuring the metal layer to add extra features and enable additional functionalities beyond the extreme localization and controlled reflectance. For instance, engineering the underlying metallic mirror to form infrared antennas  \cite{Xomalis2:2021} can lead to remarkable effects such as the detection of mid-infrared light by frequency upconversion mediated by molecular optomechanics \cite{xomalis:2021,Chen:2021}. In these cases, it would be desirable to precisely position the NPs on the underlying structured metasurface, instead of employing random drop-casting deposition. Interestingly, quite accurate positioning of NPs on planar photonic nanostructures might be obtained via a transfer process based on soft-lithography stamping \cite{Redolat:2023}.

NPs can also be deposited on narrow slits built on (functionalized) thin metal layers to form more complex photonic structures \cite{He:2019}. The resulting structure, which we have named nanoparticle-on-a-slit (NPoS) cavity (also referred to as nanoparticle-on-a-groove cavity in Ref. \cite{Chen:2021}), presents simultaneously two plasmonic gaps that double the strength of the effects resulting from light-matter interaction. In addition, the optical hot spots in the tiny region separating the metals can be easily accessed with top illumination, which is not the case in NPoM cavities, where the gap (or (1,0) mode) does not radiate in the normal direction \cite{kongsuwan:2019}. Notice that the existence of two plasmonic gaps may result in splitting the fundamental NPoM mode into two, as reported in Ref. \cite{He:2019} where vertical and horizontal bonding dipole modes were observed. However, from such modes, only the vertical one has a maximum field enhancement in the gap region and is the one we will consider in our work. Finally, it is worth suggesting that a finite slit can be considered to act as a slit antenna that resonates at a wavelength about half its length, enabling it to confine light at longer wavelengths, such as infrared. By the addition of the NP, the resulting plasmonic structure can be interpreted as a dual visible-infrared antenna \cite{Xomalis2:2021}, which is useful for optomechanical up-conversion \cite{Chen:2021}. Notably, in an NPoS cavity, deterministic positioning of the NPs on the slit would have many practical advantages over drop-casting deposition.

While metallic nanostructures can reach the ultimate limits in light localization, dielectric materials enable low-loss guidance over large distances via waveguides in photonic integrated circuits (PICs). Amongst the different technologies for photonic integration, silicon photonics \cite{Thomson:2016} has become mainstream because silicon PICs can be fabricated in high volume and at low cost using mature tools and processes from the CMOS microelectronics industry. However, the use of dielectric materials in PICs cannot lead to the extreme miniaturization provided by plasmonics as well as the large efficiency of nonlinear processes that it enables. A possible solution towards denser integration is the integration of plasmonic nanostructures on silicon PICs so that low-loss dielectric waveguides provide the guiding of light (both for excitation and collection \cite{EspinosaSoria:2016}) and light-matter-interaction processing takes place in the plasmonic elements (as in the electro-optic modulator demonstrated in Ref. \cite{Haffner:2018}), resulting in hybrid plasmonic-photonic circuits \cite{RodriguezFortuno2016}. Notably, a key benefit of such integration would be the massive multiplexing of the response of multiple on-chip plasmonic structures, which can be independently and simultaneously driven -and their response collected - by waveguides. 

A straightforward way to integrate plasmonics onto silicon PICs is by merely defining metallic nanostructures on dielectric waveguides by lithography and lift-off \cite{AlepuzBenache2012}, a strategy that has been used to observe on-chip SERS via silicon nitride waveguides \cite{Peyskens:2016,Losada2019}. The definition of metallic nanostructures by lithography has an important limitation: plasmonic gaps below 10 nm are difficult to attain due to the process resolution. Therefore, to fully exploit the advantages of nm-scale plasmonic confinement in silicon PICs, the integration of NPoM cavities with dielectric waveguide becomes highly desirable. This has been recently proposed via numerical simulations \cite{kike:2022} and demonstrated experimentally \cite{redolat:2024} but the proposed structure only works for the transverse magnetic (TM) mode of the waveguide (with the main component of the electric field pointing out of the chip plane). However, to the best of our knowledge, there have not yet been proposals or demonstrations to efficiently integrate plasmonic NPoS cavities and antennas - with the advantages over NPoM approaches that we described above - with dielectric waveguides. In addition, it would be also desirable to make it polarization- (or mode-) independent and be able to excite the plasmonic gap modes for any input polarization (or mode) of the waveguides.

In this work, we introduce a hybrid plasmonic-photonic architecture to efficiently integrate NPoS cavities/antennas with silicon nitride (SiN) waveguides in a polarization- (or mode-) independent fashion. As input and output ports, we use slot SiN waveguides connected to a thin metal plate perforated with a narrow slit that is later functionalized with a certain molecule. The SiN waveguides and the metal layer with the slit were fabricated using lithography, etching, and lift-off. Then, the Au surface is functionalized using Biphenyl-4-Thiol (BPT), which is a well-known molecule for SAM formation \cite{matei:2024}. Gold (Au) nanospheres are then carefully positioned on the slit by using a transfer process \cite{Redolat:2023} to that two nm-size plasmonic gaps per NP are created, forming the NPoS cavity. By using finite-difference time-domain simulations, we show that, counterintuitively, both TE and TM modes of the waveguide can efficiently excite the gap modes of the NPoS cavity at wavelengths around 750 nm.    
The experimental characterization of the fabricated structures, using a commercial Raman setup at an excitation wavelength of 633 nm, shows evidence of the improvement of the Raman interaction when the NP is placed on the plasmonic slit of the chip, regardless of the polarization (or mode) of the incident light. Finally, we also show that the metallic geometry behaves essentially as a patch metallic dipole antenna \cite{Zhang:2010} decorated with an NP in the gap, so it will have plasmonic resonances at the visible (due to the NPoS features) and in the infrared (by changing the length of the strips we can choose the IR resonance frequency), paving the way to new functionalities and applications of hybrid plasmonic-photonic circuits.  
 


\section*{Results}\label{sec2}
\subsection*{Description and numerical design of the on-chip hybrid structure}\label{subsec2_1}

\begin{figure*}[h!] \centering
\includegraphics[width=\textwidth]{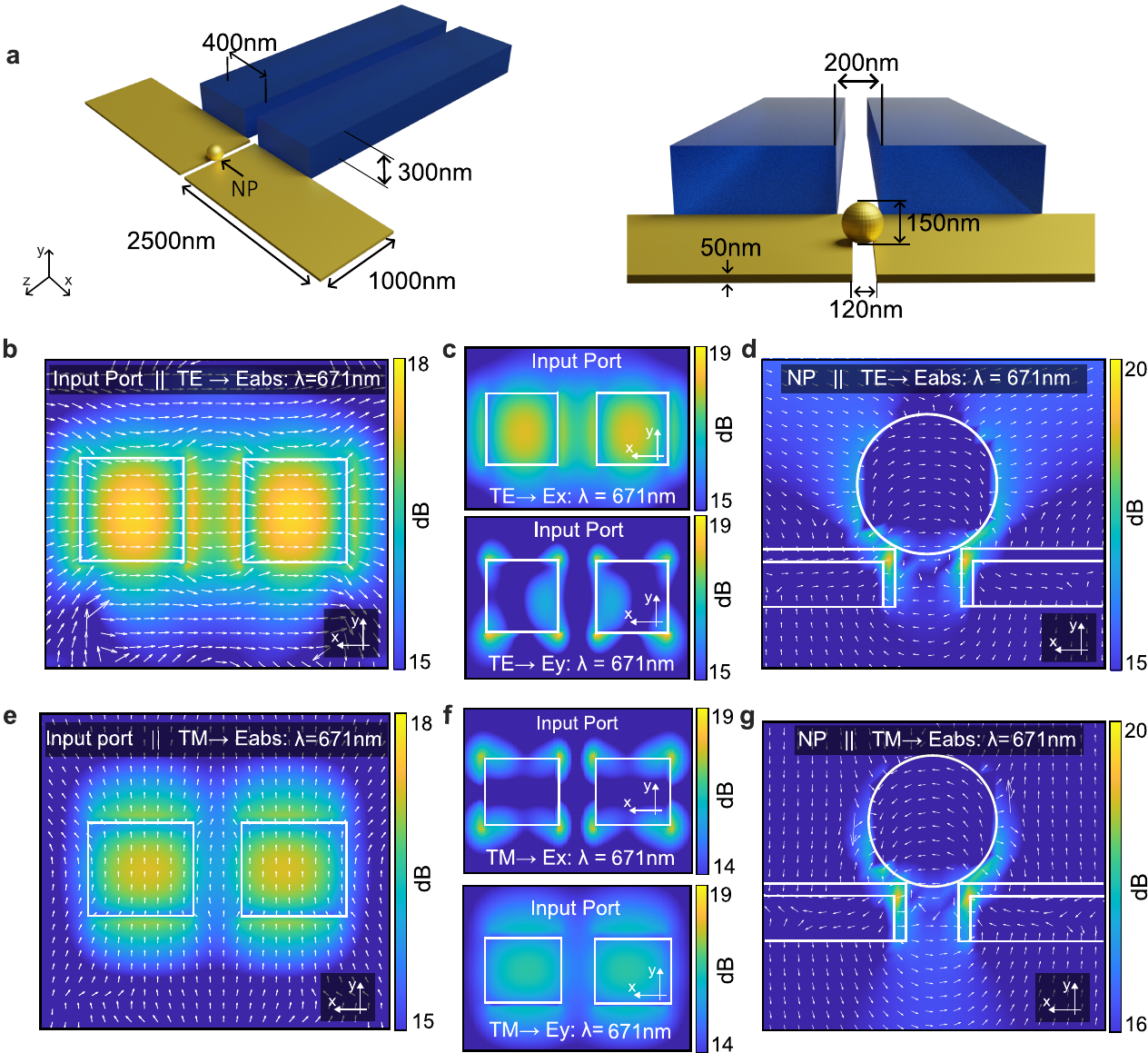}
 \caption{\textbf{Description of the plasmonic NPoS cavity integrated with SiN waveguides and illumination by guided modes.} \textbf{a} Oblique-angle (left) and frontal (right) view of a 3D sketch of the proposed configuration, including the geometrical sizes of the simulated structure. Transversal patterns of the optical signals involved in the excitation process when using the TE mode: \textbf{b}  Electric field at the excitation port, where the white arrows indicate the direction of the electric field; \textbf{c} $E_x$ (top) and $E_y$ (bottom) components of the guided field; \textbf{d} Electric field on the plane containing the nanospheres upon the metallic slit. Panels \textbf{e}, \textbf{f}, and \textbf{g} are like \textbf{b}, \textbf{c}, and \textbf{d} but for the TM-guided mode of the SiN slot waveguide. The wavelengths employed in the simulations are given in the panels. Details about the simulations are given in Methods.}
 \label{fig_1}
\end{figure*}

Figure \ref{fig_1}a shows two 3D sketches (oblique view on the left side, frontal view on the right side) of the proposed hybrid plasmonic-photonic structure. The dielectric waveguide is a SiN slot waveguide \cite{Barrios:07} placed on top of a silica substrate (not shown for clarity). Notice that SiN waveguides display negligible propagation losses at visible and near-infrared wavelengths. In the numerical simulations, we have considered a height of 300 nm and a width of 400 nm, being the two individual SiN strips separated by a 200 nm air gap. A metallic slot waveguide \cite{Veronis:2007} is placed at the end of the SiN waveguide. This plasmonic waveguide is made of two Au patches, each with dimensions of 1000 nm length, 2500 nm width, and 50 nm height, separated by a narrow slit (we assume 120 nm). Notice that this closely resembles a slit on a metallic mirror from the point of view of the creation of the plasmonic NPoS cavity. Furthermore, it can also be interpreted as a metallic dipole antenna whose resonance wavelength, which will be longer than the NPoS resonance, can be tuned by changing the length of the patches (see discussion below). An Au nanosphere, whose diameter has to be larger than the slit width (in our simulations and experiments we have considered a 150 nm diameter), is positioned on the slit at a certain distance from the SiN waveguide end. The NP is separated from the Au surface by a 1 nm thick layer with a refractive index of $n_{SAM}=$1.5, emulating the SAM functionality. The final goal of the whole architecture is that upon illumination from the waveguide the electromagnetic field is confined and intensified in the SAM below the NPs and above the Au surface, this is, in the plasmonic gaps. In the numerical simulations (see details in Methods), these points are selected as the sites for placing probes to monitor the electric field intensity and estimate the improvement in the efficiency of the Raman effect. 

For the efficient excitation of the NPoS cavity, the guided input field must include components compatible with the field localized in the plasmonic gaps, which are oblique with respect to both the $x$ and $y$ axes. In addition, this should ideally be attained for both fundamental guided modes, TE and TM, in order to achieve mode-independent excitation and collection. Notably, this is possible by using the SiN slot waveguide as the excitation (and collection) port. Although dielectric slot waveguides are mainly used to guide the TE mode that exhibits a high $E_x$ confinement in the slot (see Fig. \ref{fig_1}b), they can also propagate the TM mode, characterized by a main component of the electric field in the out-of-plane direction ($E_y$), as depicted in Fig. \ref{fig_1}e. This allows us to excite the guided modes from an external collimated beam (for instance, coming from an optical fiber) by choosing between horizontal (for TE) and vertical (for TM) polarisation. Besides the dominant component, both guided modes show an orthogonal component localized in the corners of the dielectric strips forming the slot waveguide (see Figs. \ref{fig_1}c and f) \cite{Espinosa:2016b}. Therefore, the guided-field structuring for both modes is symmetrically compatible with the field localized in the NPoS cavity gaps, as shown in the simulations depicted in Figs. \ref{fig_1}d and g. This can also be appreciated by looking at the white arrows (representing the electric field direction) in Figs. \ref{fig_1}b and e. Indeed, the local field structuring seems to give rise to the formation of a current loop around the bottom of the metallic spheres (see Figs. \ref{fig_1}d and g), mimicking a magnetic resonance that has proven to be highly efficient in terms of enhancing Raman scattering \cite{Chen:2018}. 

\begin{figure*}[h!] \centering
 \includegraphics[width=\textwidth]{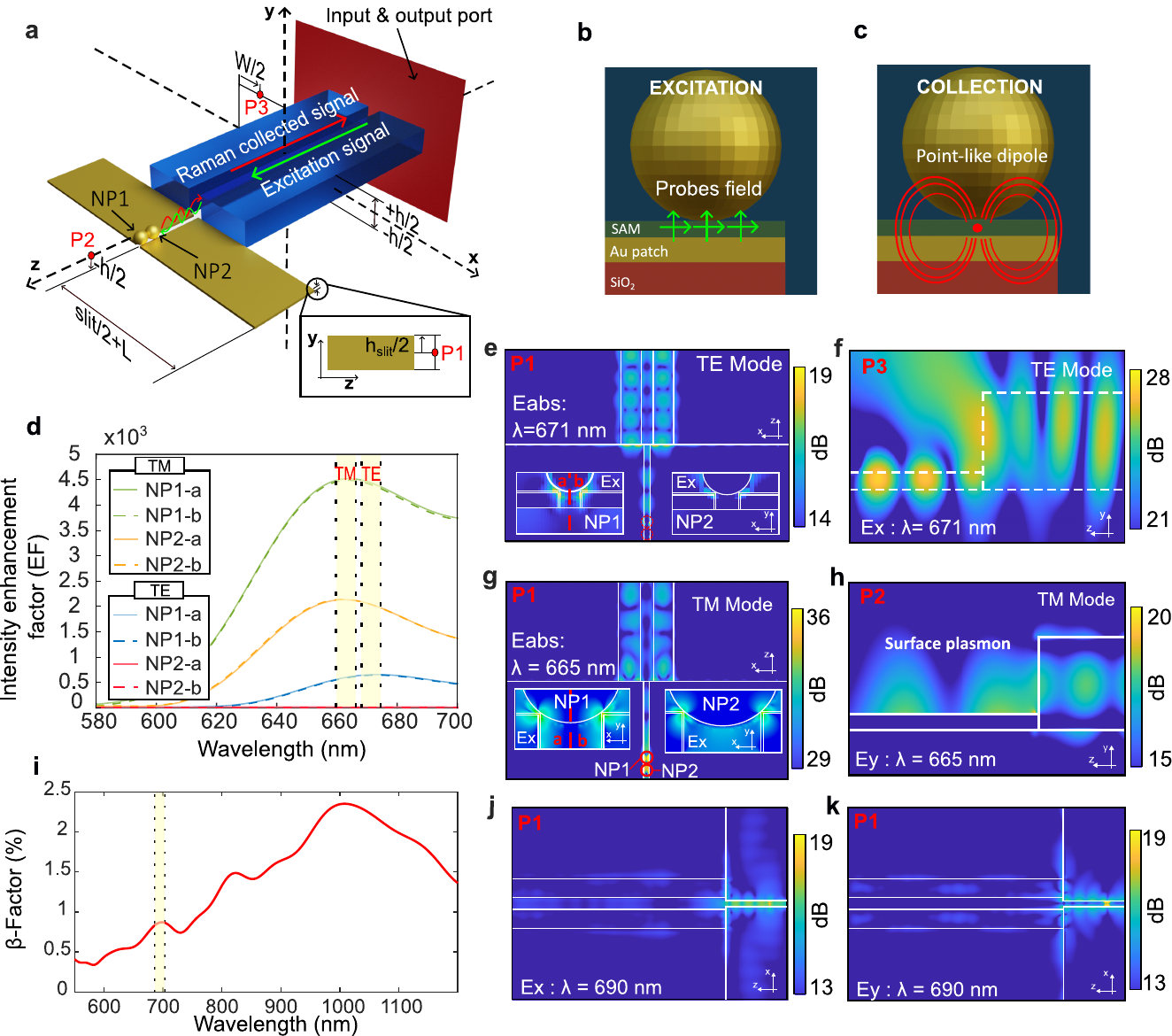}
 \caption{\textbf{Numerical results of the excitation of the NPoS cavity using SiN slot waveguides.} \textbf{a} 3D scheme of the configuration used in the numerical simulations, where the red square represents the input and output port, the green arrows correspond to the excitation signal and the red arrow represents the backscattered Raman signal. \textbf{b} Scheme of the probe locations for the study of the field enhancement in the excitation simulations. \textbf{c} Representation of the point-like dipole source emulating a Raman center used in the evaluation of the $\beta-$factor in the collection scheme. \textbf{d} Enhancement factor (EF) in the central probe field in \textbf{d} for both the TE and TM modes as a function of the wavelength. \textbf{e} Top-view and \textbf{f} side-view snapshots of the propagation of the $E_{abs}$ absolute of the TE mode for $\lambda = 671 nm$. The insets in \textbf{e} show the transversal field maps under the same illumination conditions in the positions of NP1 (left) and NP2 (right) $E_x$ component. \textbf{g} Top-view snapshot of the absolute electric field $E_{abs}$ and \textbf{h} side-view snapshot of the the $E_y$ component when illuminating with the TM guided mode at $\lambda = 665 nm$. The insets in \textbf{g} show the transversal field maps under the same illumination conditions in the positions of NP1 (left) and NP2 (right) of the $E_x$ component. The inset in \textbf{g} shows that the TM mode excites a surface plasmon on each metal film forming the Au slot waveguide.  \textbf{i} Simulated  $\beta-$factor as a function of the wavelength when placing a $y$-oriented point-like dipole in the position shown in \textbf{c} for NP1. Top-view snapshots of the \textbf{j} $E_x$ and \textbf{k} $E_y$ electric field components when the input to the system is a point-like dipole at $\lambda = 690 nm$.}
 \label{fig_2}
\end{figure*}

The performance of the whole hybrid plasmonic-photonic device was simulated using a commercial electrodynamics solver (CST Studio Suite). We considered the geometry depicted in Fig. \ref{fig_2}a using the same plane as an entrance and exit signal port (red square in Fig. \ref{fig_2}a). In these simulations, we included two NPs on the slit to mimic the sample that was fabricated and experimentally measured. Essentially, the study was conducted by assessing the performance in two distinct situations: excitation and collection \cite{Losada2019}. On the one hand, the excitation efficiency was evaluated as the intensity enhancement factor (EF), defined as $EF(r,\lambda)\equiv|E(r,\lambda)|^{2}/|E_{0}(r,\lambda)|^{2}$, where $E(r,\lambda)$ correspond to the field measured at a certain position $r$ considering the presence of the whole metallic structure and $E_{0}(r,\lambda)$ is the field resulting at the same position but without the resonator structure (this is, by completely removing the metal). To estimate the enhancement of the excitation efficiency of the Raman centers in the plasmonic gaps, we placed a set of field probes in the places depicted in Fig. \ref{fig_2}b (this is, in the middle of the SAM separating the NP and the slit corners) and recorded the local field when a guided mode illuminates the whole system. The three probes, which were spaced by 1/4 of the NP diameter, were duplicated at each side of the waveguide (in the panels of Fig.\ref{fig_2}, we use the subscripts $a$ and $b$ to refer to the probes on the left and the right side of the waveguide, respectively) to account for the two plasmonic gaps per NP. On the other hand, the collection efficiency was evaluated by utilizing the $\beta-$factor, defined as $\beta= P_{TMode}/P_{Raman}$, where $P_{TMode}$ denotes the optical power that couples into the guided modes of the input waveguide, and $P_{Raman}$ represents the overall power released by a point-like source acting as a Raman scatterer. In our case, we considered a $y$-axis orientation, and it was located in the middle of the SAM, at the same position as the middle probe point in the excitation configuration (see Fig. \ref{fig_2}c).

Figure \ref{fig_2}d shows the intensity enhancement EF as a function of the wavelength (see simulation details in Methods), where we can see how the field intensity is enhanced up to $\approx$ 4.5x$10^{3}$ for the TM mode and $\approx$ 6500 for the TE mode. Such EF maxima occur at the wavelengths of $\lambda = 644 nm$ and $\lambda = 665 nm$ for the TM and TE modes, respectively, which are within the plasmonic resonance of the NPoS cavity (approximately covering wavelengths between 550 nm and 700 nm). Although these values may not seem too high, we must stress that the normalization is carried out with respect to the system without the metal, where there is no plasmonic localization but still there is wavelength-size confinement in the integrated slot waveguide. The enhancement spectrum confirms the suppositions we made during the analysis related to Fig. \ref{fig_1}: with our proposed structure, the plasmonic gaps can be excited using both the TE and TM guided modes. This can also be confirmed using continuous wave numerical simulations at the wavelengths at which the enhancement reaches a maximum for either the TE or the TM mode. To visualize the excitation of the plasmonic gaps, Fig. \ref{fig_2}e  shows a snapshot of the absolute electric field ($\lambda = 665 nm$) along the slot waveguide for TE mode excitation and how it couples first to the plasmonic slot guide and later to both plasmonic gaps between the NP and the Au layer.  In the insets of Fig. \ref{fig_2}e, we can see that the (1,0) cavity mode is excited for both NPs (additional information in Supplementary Fig. 3), although the intensity of the enhanced field for NP1 is stronger than NP2, which is expected since some light is lost (due to both absorption and scattering) during the interaction with NP1. Similar results are obtained for excitation with the TM mode ( though now at $\lambda = 644nm$: as shown in the insets of Fig. \ref{fig_2}g, there is a huge confinement in the gaps, which leads us to conclude that we are exciting plasmonic modes spatially mimicking the (1,0) mode of the NPoM cavity (the vertical bonding dipole in \cite{He:2019}) ), as in the case of the TE excitation. We also show lateral snapshots of the propagating field for both modes (see Fig. \ref{fig_2}f and h). 
Noticeably, when using the TE mode for illumination, we excite the slot mode of the metallic slit, followed by the (1,0) modes of the cavity. In contrast, when using the TM mode of the SiN waveguide, propagation in the metallic region occurs as two surface plasmons, each on a metallic patch. These plasmons are excited by the field individually guided by each of the waveguides forming the dielectric slot and can also efficiently couple to the (1,0) cavity mode. Notice that the field in the gap vanished for the (or vertical bonding dipole \cite{He:2019} of the NPoS cavity, so the excitation of different plasmonic modes does not seem the right explanation for the slightly different behaviour when exciting with either the TE or the TM mode. Therefore, the observed difference in the EF between the cases of TE or TM excitation is most probably caused by the different structuring of the propagation modes of the optical field in the metallic slot as well as a different coupling efficiency between the dielectric and the metallic guided modes.

Figure \ref{fig_2}i depicts the calculated $\beta-$factor as a function of the wavelength under the assumption that Raman centers are stimulated in the gaps of both NPs. A collection efficiency of $\beta$ $\approx$ 1\% is observed when placing a dipole-like source mimicking a Raman emitter placed in one of the gaps of NP1 in the wavelength region around 690 nm that is close to the EF maxima observed in the excitation spectra of Fig. \ref{fig_2}d. Notice that, in a real experiment, there will be multiple Raman sources contributing to the output signals, all of them excited by the same input signal. Remarkably, higher values of the $\beta-$factor (peaks reaching 2.5\% at wavelengths around 1 $\mu$m) are observed in the spectrum at longer wavelengths. The reason is unknown, but it can be related to the fact that the dimensions of the SiN waveguides, with a thickness of 300 nm, are more appropriate for guiding near-infrared light in the telecom regime \cite{CastellPedrero:2021} and therefore more signal is collected at those longer wavelengths. To visualize the capture of the scattered Raman signal by the waveguide, Figs. \ref{fig_2}j and k depict simulated snapshots of the horizontal $E_x$ and vertical $E_y$ electric field components when a dipole tuned at $\lambda = 690 nm$ is positioned in a plasmonic gap. This illustrates the coupling of the emitted radiation to the two waveguide modes: the $E_x$ component corresponds to an output TE mode and the $E_y$ component to the TM one. Notice that the whole structure has not been optimized either for excitation or collection, but it is realistic to think that the obtained values can be further improved by a careful design of the whole system.


\newpage
\subsection*{Experimental verification}\label{subsec2_2}

\begin{figure}[H]\centering
 \includegraphics[width=\textwidth]{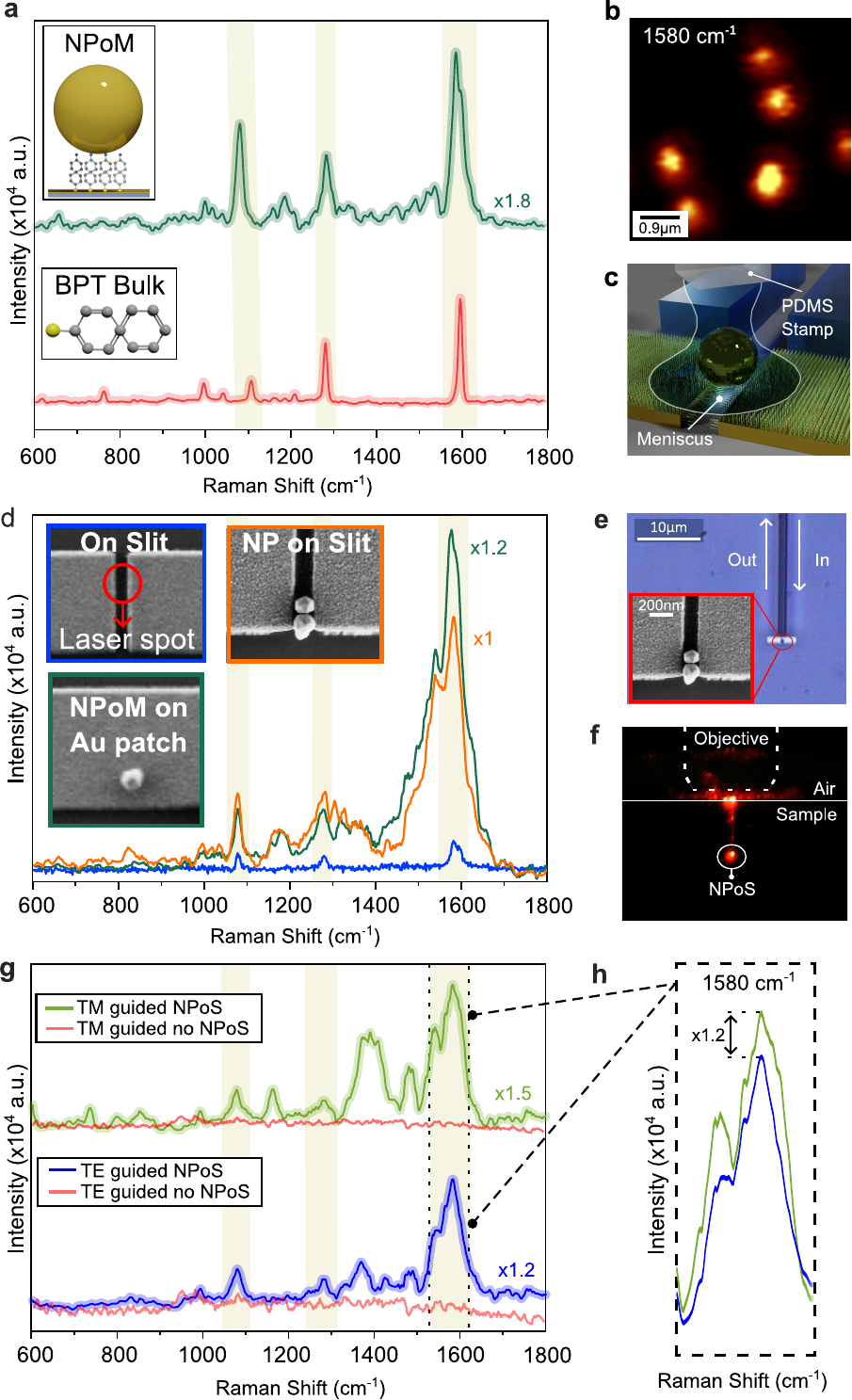}
 \caption{\textbf{Experimental Raman characterization of the fabricated samples}.\textbf{a} Characteristic Raman spectrum of BPT on a metal surface without structuring (red line) and after forming NPoM cavities (green line). \textbf{b}  Raman image of 1580 $cm^{-1}$ BPT Raman peak of the Au substrate functionalized with BPT and some NPoM created by drop casting. \textbf{c} Scheme of how the NPs are transferred in the slit antenna by the formation of a meniscus when using the soft lithography method. \textbf{d} BPT Raman spectrum for $\lambda$ = 633nm laser exciting and collecting from the top comparing all possibilities of the functionalized Au antenna: only the antenna (blue line), NPoM on the antenna Au patch surface (green line) and NPoS (orange line). Additionally, SEM images of all alternatives are presented as insets. The color of the SEM image edges corresponds to the color spectra lines. \textbf{e} Optical image of the fabricated photonic component after NP positioning where the white arrows indicate the path for the incident and collected light. The inset shows an SEM image of the NPoS structure. \textbf{f} Optical top-scattering image showing how the light is coupled in the slot waveguide directing it to the NPoS cavity, which is excited. \textbf{g} Guided detected Raman of BPT when the nanocavity is excited with TE polarization (blue line) and TM polarization (green line). The Raman spectra obtained from the functionalized samples before the NPs transfer are also shown for comparison for both input modes. \textbf{h} Zoom-in of the 1580 $cm^{-1}$ BPT Raman peak comparing the recorded intensities for TE (blue line) and TM mode (green line) mode driving.}
 \label{fig_3}
\end{figure}

The hybrid plasmonic-photonic structure was fabricated using a synergistic combination of top-down and bottom-up lithography techniques. The dielectric slot waveguide was patterned on a standard, 300 nm-thickness SiN wafer via electron beam lithography and a reactive ion etching. After a second electron beam exposure, the two thin Au layers forming the metallic slit waveguide (or dipole antenna) were evaporated at the waveguide end before using lift-off to remove the residual resist. A thin (5 nm) Chromium layer was evaporated before Au to improve its adhesion to the silica substrate. Then, the Au layers were functionalized with BPT, a process in which the molecule adheres to the metallic surface through the thiol interaction. We characterized this process separately by synthesizing the SAM on an 80 nm-thick Au layer on silicon without nanostructuring (see details in Methods and also Supplementary Fig. 1). We used advancing-receding water contact angle measurements to verify the presence of the SAM on the Au surfaces, as shown in Supplementary Fig.2 (see also Methods). Then we used Raman spectroscopy to test the properties of the SAM. Notice that we adequately subtracted the background in all Raman spectra represented in this work. When recording the Raman spectrum from the functionalized Au layer, a set of scattering peaks was clearly observed (see the red curve in Fig. \ref{fig_3}a). The main peaks (1090, 1280, and 1580 $cm^{-1}$) can be attributed to the vibrational resonances of the BPT molecule, in good agreement with previous experiments \cite{baumberg:2019, chikkaraddy:2022,  xomalis:2021, benz:2016, lombardi:2018}. These peaks also match the expected values according to a theoretical study \cite{rosta:2022} (see Supplementary Section 1). Then we carried out drop-casting of Au NPs on the same sample to form NPoM cavities that enhanced the Raman interaction. The resulting Raman spectra show a net increase in the recorded Raman signal (see green curve in Fig. \ref{fig_3}a). Indeed, we performed a Raman mapping on the surface (see Fig. \ref{fig_3}b) that clearly shows the huge enhancement of the BPT Raman peak at 1580 $cm^{-1}$ in the positions of the NPs.

Then we proceeded with the controlled positioning of Au NPs to the fabricated sample containing the SiN slot waveguides as well as the plasmonic slit covered with the SAM according to the processes explained above. The NP transfer process is schematically depicted in Fig. \ref{fig_3}c. Notice that even though this transfer method should enable the positioning of single NPs on photonic structures with high accuracy \cite{Redolat:2023}, the plasmonic slit is so small that the whole process becomes highly challenging. Still, for the best device, we achieved positioning two NPs (which were both trapped in the meniscus formed in the polymeric stamp during the soft lithography process) in the slit, therefore forming a dual-particle NPoS cavity, as shown in Figs. \ref{fig_3}d and e. However, in another plasmonic cavity fabricated on the same sample, a single NP was transferred to one of the metallic patches instead of on the slit (see bottom inset in Fig. \ref{fig_3}d). We measured the Raman spectra of the three different plasmonic structures depicted in the insets of Fig. \ref{fig_3}d using free-space light incident normally to the sample. The results, shown in Fig. \ref{fig_3}d, show that the Raman signatures of the BPT molecules are present in our samples, being the Raman response largely enhanced when the NPs are present either in the slit (x100) or on top (x120) of one of the metal patches. Notice that accurate nanopositioning of the NPs in the slits is also feasible using dielectrophoresis trapping \cite{He:2019}. This approach, though also feasible in PICs, would make the whole device more complex since it requires electrical contacts that should not interfere with the optical guided waves to avoid extra losses. 


To perform the Raman measurements with guided light, we adapted the microscope of our Raman spectrometer (more details in Methods and also Supplementary Fig. 4). This adaptation allowed us to change the polarization of the illumination signal to excite either the TM or the TE mode of the input waveguide. Additionally, we were able to capture top images of the scattered light during the measurement process. Figure \ref{fig_3}f  shows how the light is coupled to the slot waveguide and directed towards the NPoS cavity, in good agreement with the simulations shown in Fig. \ref{fig_2}. Note that no polarization filtering was applied to the Raman collected signal so we were not able to discern between TE and TM modes in collection. Figure \ref{fig_3}g presents the obtained Raman spectra (after background subtraction) when using guided light to both excite the molecules and collect the Raman scattering. In both spectra, we can see the resonance peaks corresponding to the vibrational modes of the BPT molecule (1090, 1280, and 1580 $cm^{-1}$) consistent with our previous measurements using free-space excitation and collection. Notably, the BPT molecules were excited and detected for both TE mode (blue line in Fig. \ref{fig_3}g) and TM mode (green line in Fig. \ref{fig_3}g) but only after transfer of the NPs. No appreciable Raman peaks are observed in the guided measurements of the plasmonic slit without NPs. Furthermore, we can observe that the signal corresponding to the TM mode is more intense than that for the TE mode (see the zoomed 1580 $cm^{-1}$ peak in Fig. \ref{fig_3}h), validating the results obtained with numerical simulations. In addition to the use of different input polarizations/modes with distinct efficiencies, the slight differences in both spectra could be due to what is known as the selection rule, which determines that the orientation of the molecule relative to the incident beam can excite an additional vibrational mode \cite{Moskovits:1980, Moskovits:1982, xioping:1990}. 

\begin{figure}[H]\centering
 \includegraphics[width=\textwidth]{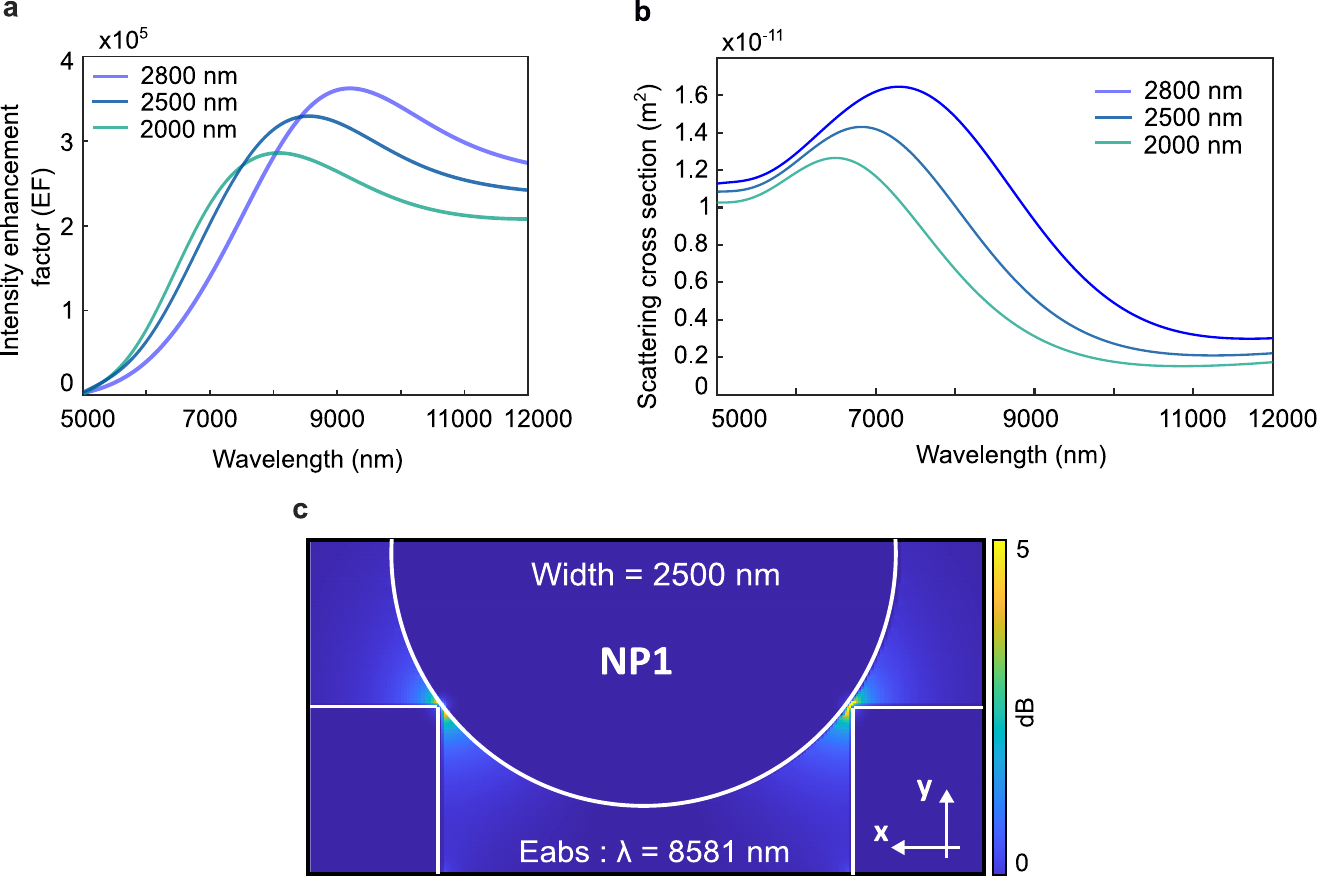}
 \caption{\textbf{Numerical modeling of the NPoS antenna at infrared wavelength range}.\textbf{a.} Maximum of the E field intensity enhancement for different sizes of the antenna at infrared wavelengths. \textbf{b.} Scattering cross-section spectra for different sizes of the antenna. \textbf{c.} Transversal field maps view of the confined and enhanced field in the gaps of NP1 when the structure is illuminated by a normally-incident plane wave at $\lambda = 8581 nm$.}
 \label{fig_4}
\end{figure}

As mentioned above, our plasmonic cavity can also be viewed as a metallic-patch dipole antenna decorated with an NP in the gap, as can be appreciated in Fig. \ref{fig_1}a. When illuminated from the top, such an antenna will present an optical resonance at a wavelength about twice the total length. Such resonance will be characterized by a huge field concentration in the slit, which in principle should be further intensified and fully localized in the nm-size gaps by the presence of the NPs. We have performed FDTD simulations (see Methods) to calculate the scattering cross-section as well as the field intensity enhancement for normal illumination. As shown in Figs. \ref{fig_4}a and b, the antenna displays a resonant response at infrared wavelengths for the selected values of the antenna length. Moreover, the resonance is shifted by changing the length of the antenna, as expected. In order to visualize the confinement of the electromagnetic radiation in the small gap between the NP and the antenna, we have represented the near-field maps at the wavelength where the scattering cross-section is maximum (see Fig. \ref{fig_4}c). It can be seen that infrared light is tightly confined in the gaps, as it also happens with visible radiation for guided illumination. This confirms the dual-wavelength performance of the hybrid structure, which could be employed for on-chip infrared (or even terahertz \cite{Roelli:2020}) detection mediated by molecular optomechanics \cite{Chen:2021,Xomalis2:2021}.

\section*{Conclusions}\label{sec4}
In summary, we have introduced and demonstrated a hybrid plasmonic-photonic configuration with interesting properties for on-chip waveguided Raman spectroscopy \cite{ettabib:2021, ettabib:2024} and THz/infrared detection via molecular optomechanics \cite{Roelli:2020}. The structure consists of a dielectric slot waveguide as an input and output port that is connected at its end to a metallic slit waveguide covered by a SAM and on which metallic NPs are deposited using nanoimprint transfer. This plasmonic cavity, which we call nanoparticle-on-a-slit, constitutes a variation of the traditional NPoM cavity but inherits its main feature: the possibility to localize the optical field beyond the diffraction limit in a volume orders of magnitude smaller than a cubic wavelength. In addition, we have shown that the cavity has a dual-wavelength behavior and can collect visible light coming from the waveguides and infrared/THz light when illuminated from the top. In this last case, the metallic structure essentially behaves as a dipole antenna decorated by an NP in its gap.

Unlike other hybrid structures integrating plasmonic cavities and dielectric waveguides \cite{Peyskens:2016,Losada2019}, here the plasmonic gap is not defined by lithography but by self-assembly, which enables a much narrower plasmonic gap (and therefore, a higher field enhancement) only limited by the thickness of the deposited monolayer. Therefore, it can contribute to completing the landscape of available hybrid plasmonic-photonic circuits \cite{Haffner:2018} with the added value of achieving nm-size plasmonic gaps. The demonstrated structure also possesses several advantages with respect to a previous approach to integrating dielectric waveguides with NPoM cavities \cite{redolat:2024}, in particular, the possibility to drive the cavity with either TE or TM guided modes (this is, the device is polarization blind in the sense) and the existence of two optical hot-spots per NP, which can result in a larger collection of the Raman signal. Although here demonstrated for BPT, other molecules could be self-assembled over the Au surfaces before transferring the NPs, opening the path for new effects. In addition, we highlight that there is room for improvement: by properly optimizing several parameters, it should be possible to achieve larger efficiencies in the transduction of the Raman signal, ultimately leading to the recording of Raman spectra of single molecules in an on-chip approach. Although not observed in our experiments, our structure is also adequate for the observation of picocavities \cite{benz:2016} using guided light in a silicon photonic integrated circuit.

\section*{Materials and methods}\label{sec4}
\subsection*{Simulations}\label{subsec3_1}
The analytical study was conducted using two simulation tools, \textit{CST Microwave Studio} and \textit{Lumerical}, for the visible range (using the Finite Element Method (FEM)) and for the infrared range (using the Finite-Difference Time-Domain (FDTD) method), respectively. The geometry studied was the same in both cases. The SiN waveguides were simulated as two bars with a width (dimension along the $y$-axis, see Fig. \ref{fig_1}a) of 400 nm and a height (dimension along the $x$-axis, see Fig. \ref{fig_1}a) of 300 nm. The gap between the waveguides was 200 nm. The optical properties of SiN were considered constant within the analyzed spectral range ($n_{SiN} = 2$). The waveguides were placed on a semi-infinite silica substrate with a refractive index $n_s=1.5$. The Au dipole antenna was positioned transverse to the SiN waveguides. To design the metallic antenna, two Au identical patches of 2500 nm width, separated by a 120 nm gap, and with a length (dimension along the $x$-axis, see Fig. \ref{fig_1}a in the manuscript) of 1000 nm and a height (dimension along the $y$-axis, see Fig. \ref{fig_1}a in the manuscript) of 50 nm were created. The optical properties of gold were modeled using a Drude model. A gold sphere with a radius of 75 nm was placed in the slit of the nanoantenna. The position of the nanoparticle was such that it was constrained within the walls of the metallic slit.

Regarding the simulation strategy for guided waves in the visible regime, two parallel sets of simulations were performed. On one hand, excitation studies were conducted by placing field probes to obtain the electric field intensity just below the NP and at the midpoint of the gap between the NP and the Au patch, numbered in the direction of light propagation along the $z$-axis (as shown in the inset of Fig. 2a). The intensities were measured with and without the presence of the metallic structure to normalize the signal and obtain the enhancement. The signal coupled to the waveguides came from a port placed at the opposite end of the antenna, at the beginning of the slot waveguide (red square of Fig. \ref{fig_2}a). On the other hand, to study the collection, a point-like dipole was placed in the same position as the central probe (probe number 2 in Fig. \ref{fig_2}c). In this case, the S-parameters of the port placed at the beginning of the slot waveguide indicated the amount of signal from the dipole that was coupled back into the waveguide and transported to the other end, where the port was located. To analyze the electric field propagation, field monitors were placed for the wavelengths where the enhancement is highest, in the case of excitation, and the wavelength corresponding to the highest collection peak in the visible range. The different obtained planes can be seen in Fig. \ref{fig_2}a: the yz plane at the height of point P2 (a symmetrical cut in the center of the gap), the $yz$-plane at the height of point P3 (dividing a waveguide symmetrically), and the $xz$-plane through point P2 dividing the antenna into two parts for an overhead view. Additionally, cuts in the $yz$-plane that divide each NP into two equal parts are shown, and a final cut to visualize how TM and TE modes couple in the slot waveguide. The mesh used was hexahedral and was individually refined for each structure, making it more precise in the region where the electric field confinement occurs.

Regarding the infrared simulations, the FDTD simulation region was a cube with a side length of 7 $\mu$m. Perfectly matched layers (PMLs) were used to absorb the scattered field. The structure was illuminated by a total-field scattered-field (TFSF) source, which launched a broadband plane wave from the top of the system under normal incidence. To calculate the scattering cross-section, an analysis group consisting of power monitors outside the source was employed. The scattered power was determined by integrating the Poynting vector of the scattered field over the monitors, which formed a box surrounding the source. The scattering cross-section was then obtained by dividing the scattered power by the incident intensity, similar to the visible range simulations. Near-field maps were generated using a frequency-domain field and power monitor in the $XY$-plane, at the midpoint of the Au nanoparticle. To ensure full convergence of the solutions, automatic non-uniform meshes were utilized with a 0.7 nm mesh refinement along the $x$, $y$, and $z$-axes in the region around the nanoslit where the gold nanoparticle is located.

\subsubsection*{Sample functionalization}\label{subsec3_2}
Biphenyl-4-thiol (BPT, 97$\%$) molecules were acquired from Merck-Sigma Aldrich. Initially, a piranha solution (H2SO4/H2O2, 1:1) was employed to clean glass surfaces. The BPT self-assembled monolayers (SAMs) were then created by immersing the substrates in a 1 mM solution of BPT in absolute ethanol for 14 hours. Subsequently, the samples were sonicated in ethanol for 3 minutes, washed with ethanol, and dried under a stream of nitrogen $N_2$ gas.

\subsubsection*{Nanoparticle transfer}\label{subsec3_3}
A soft lithography method was employed to place the Au NPs on the slit \cite{Redolat:2023}. The setup offers a three-axis movement and rotation for the alignment with the sample. Spherical citrate-capped 150nm Au-NPs (from Nanopartz$^{TM}$) were accurately positioned on a gap of 130nm between two Au patches to create the NPoS cavity. Previously the transfer, the Au surfaces were functionalized. Finally, the antenna was integrated with a slot waveguide for illumination and collection in a guided way. For drop casting, the NPs were delivered onto the surface for 2$'$, then the sample was cleaned with deionized water and dried with $N_2$.

\subsubsection*{Raman measurements}\label{subsec3_4}
Raman images and Raman spectral information were acquired using a Spectrometer alpha300 RA (Raman-AFM) from WITec, providing information about the Antistokes Raman scattering in the 80-3000 cm$^{-1}$ range. A 633nm laser illumination system and 100x objective complete the equipment. A laser power $P$ = 10 mW and a grating spacing $G$=600 l/mm were the used parameters. For single spectra, an integration time of 0.1s was used. For the Raman images, the sample was scanned with a 0.035s integration time and using 30x30 points per line image.

\subsubsection*{Contact angle}\label{subsec3_5}
The quality of the SAMs was estimated using advancing and receding contact angle measurements with a Ramé-hart automated goniometer. WCA measurements of both functionalized and non-functionalized samples were conducted using the needle-in-sessile-drop technique at room temperature, without humidity control. Initially, a deionized water droplet of approximately 1mL was dropped onto the surface of the sample using a motorized micro syringe. Small amounts of water (1µL increments) were then added to the droplet in single steps while CA was recorded. Then, the process was repeated, removing, in single steps, the same amount of volume each. In total, the whole process consists of three times repetitions of the same cycle (10 steps adding volume + 10 steps removing volume). The average of the highest (lowest) angles from the last four cycles of the advancing (receding) contact angles were then calculated.

\subsubsection*{Atomic force microscope}\label{subsec3_6}
An Alpha300 RA (Raman-AFM) instrument from WITec was employed for the AFM sample characterization. All measurements were performed in AC mode. Sharp silicon 4 probes without coating (K ~ 42 N/m, f0 ~ 320 kHz) were purchased from PPP-NCH (Nanosensors). All AFM images were processed with WSxM software from Nanotec Electrónica S.L.1

\subsubsection*{Scannig Electron Microscopy imaging}\label{subsec7}
High-resolution field emission scanning electron microscopy (HRFESEM) was employed for the imaging of the fabricated structures. The configuration of the HRFESEM was using a ZEISS GeminiSEM 500 with 0.5nm at 15kV, 0.9 nm at 1kV, and 1nm at 500V resolution.

\subsection*{Sample fabrication}\label{sup_fab} 
The waveguide structures were produced on standard silicon nitride substrates with a thickness of 300 nm and a buried oxide layer of 3.26 µm. The production involved an electron-beam (e-beam) direct-writing method on a 300 nm thick negative (Man-2403) resist coating. This e-beam exposure, conducted using a Raith150 tool, was finely tuned to achieve the desired dimensions, using an acceleration voltage of 20 KeV and an aperture size of 30 µm. After the development phase, the resist patterns were etched into the silicon nitride using a refined Inductively Coupled Plasma-Reactive Ion Etching process with fluoride gases. A subsequent e-beam lithography step using positive resist PMMA was executed to create the metal patches before a 40 nm gold evaporation and a lift-off process using MNP as the solvent.




\backmatter

\section*{Data availability}
Data supporting this research will be made available in an open repository.

\section*{Acknowledgements}
We acknowledge funding from the European Commission (THOR project, Grant agreement No. 829067, and 'NextGenerationEU'/PRTR and 'ERDF A way of making Europe'), Generlaitat Valenciana under grant CIPROM/2022/14, and the Spanish Ministry of Science and Innovation (MCIN/AEI/10.13039/501100011033) under grant PID2021-124618NB-C21. E.P.C gratefully acknowledges funding from Generalitat Valenciana (Grant No. SEJIGENT/2021/039), AGENCIA ESTATAL DE INVESTIGACIÓN of Ministerio de Ciencia e Innovación (PID2021-128442NA-I00) and the European Regional Development Fund (ERDF). A.B. thanks MICINN for the Ramón y Cajal Fellowship (grant No. RYC2021-030880-I) and the Spanish Research Agency (AEI) for the project No. PID2022-137857NA-100. J.R. Acknowledges funding from Universitat Politècnica de València (Grant No. FPI 20-10253).


\section*{Author contributions}
J.R. conducted the experimental measurements including
Raman spectroscopy, surface functionalization, NP transfer, and analyzed Raman spectra. D.A-O. helped with experimental measurements and the NP transfer. A. B. and J.R. performed the numerical simulations of the system. A.G. fabricated the dielectric and plasmonic waveguides. A.M. conceived the idea of the NPos integration with dielectric slot waveguides, and supervised the experiments and numerical simulations. E.P.-C supervised
the sample fabrication, the NP imprinting, and the Raman measurements, and
analyzed the Raman spectra. The manuscript was written through the contributions of all authors. All authors have approved the final version of the manuscript

\section*{Competing interests}
The authors declare no competing interests.


\begin{appendices}
\section{Supplementary information}\label{secA1}

\section*{S1. BPT molecule}

The 1-1 biphenyl 4-thiol (BPT) molecule used in our experiments consists of two phenyl rings connected at the first carbon of each ring, with a thiol group (-SH) attached to the fourth carbon of one phenyl ring, and has the molecular formula $C_{12}H_{10}S$. This structure includes an aromatic biphenyl system and a reactive thiol group, making it appropriate to form self-assembled monolayers (SAM) on metal surfaces (see the structure in Supplementary Fig. \ref{sup_molecule}a). The theoretical Raman spectrum of BPT, shown in Supplementary Fig. \ref{sup_molecule}b, is a good reference to check the BPT resonances obtained in our experiments (see main text).

\begin{figure}[h!]
 \centering
 \includegraphics[width=0.5\textwidth]{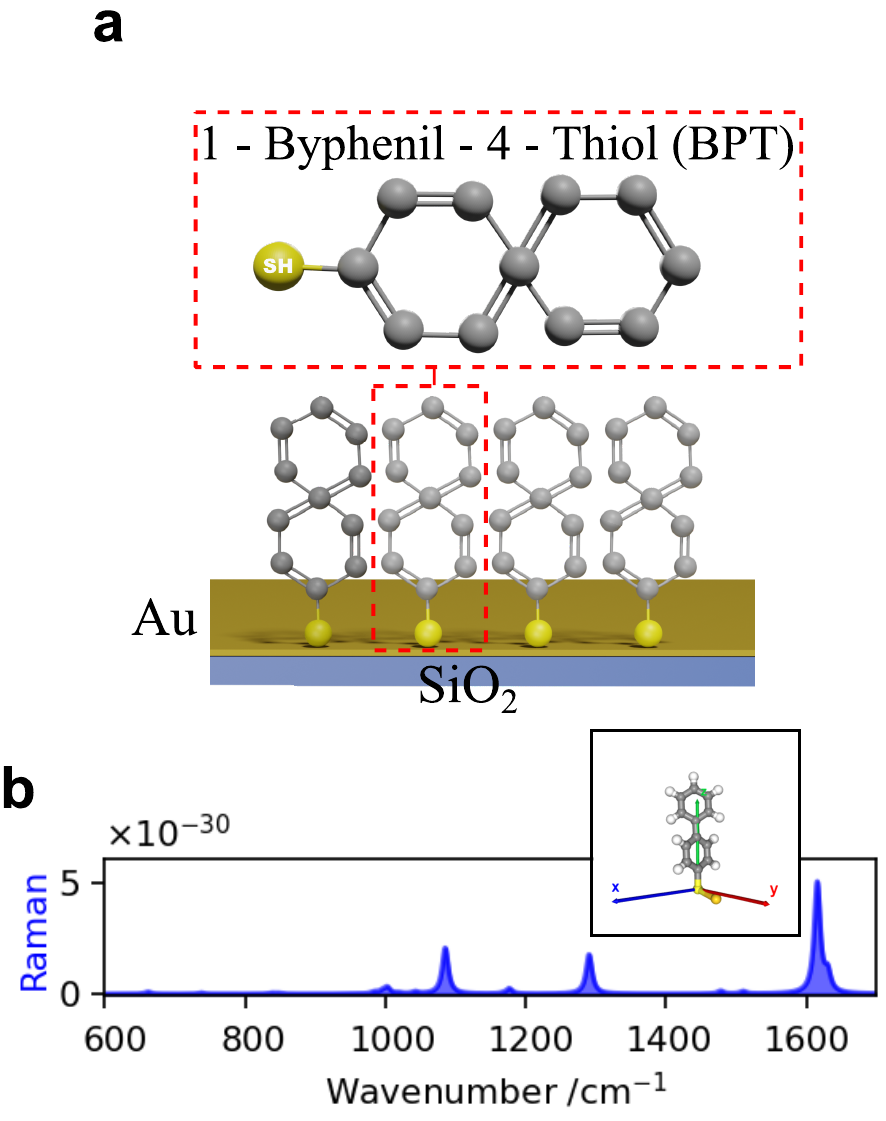}
 \caption{\textbf{BPT molecule, SAM formation and simulated Raman response.} \textbf{a} BPT molecule and schematic SAM structure formed on an Au surface. \textbf{b} Raman spectra obtained by simulations at the range of 15-60THz (500-1700$cm^{-1}$), considering a 633nm Raman laser, x-polarization, and ambient conditions of BPT molecules from ref. \cite{rosta:2022}. Inset: BPT molecule orientation. SMILES code: $c1ccc(-c2cccc2)cc1$. Note on spectra units: THz/IR $[K \cdot mol^{-1}]$, Raman $[cm^2 \cdot sr^{-1}]$, conversion $[Km \cdot mol^{-1} \cdot 1 \cdot cm^{-2} \cdot^{-1}]$.}
 \label{sup_molecule}
\end{figure}

\section*{S2. Contact angle and atomic force microscopy (AFM)}\label{sup_s2} 
These measurements were performed with Ramé-hart Model 90 Standard Goniometer and Dropimage Standard software. An automated dispensing system is installed to control the drop volume during the advancing and receiving contact angle method. This setup integrates software, an LED illuminator, a 3-axis levelling stage, a digital camera, and a microsyringe fixture and assembly for manual dispensing. The dispensing system and a manual tilting base contribute to increasing the data acquisition precision.

Two samples were used to measure the changes between a non-functionalized and BPT-functionalized gold surface. The wate contact angle (CA) results are shown in supplementary Fig.\ref{fig_ca_supp}a, where a difference of around $10^{\circ}$ in a non-functionalized (blue line) and functionalized (red line) is clearly appreciable. The differences in the (CA) values allow us to confirm the presence of the molecule on the surface. CA results are $85^{\circ}$ and $73^{\circ}$ for a non-functionalized sample and a functionalized one, respectively. The bare surface was clearly with ethanol absolute and dried with $N_2$ before the quantification of the CA. The functionalization was made following the steps explained in the Materials and methods (main text).

\begin{figure}[H]
 \includegraphics[width=\textwidth]{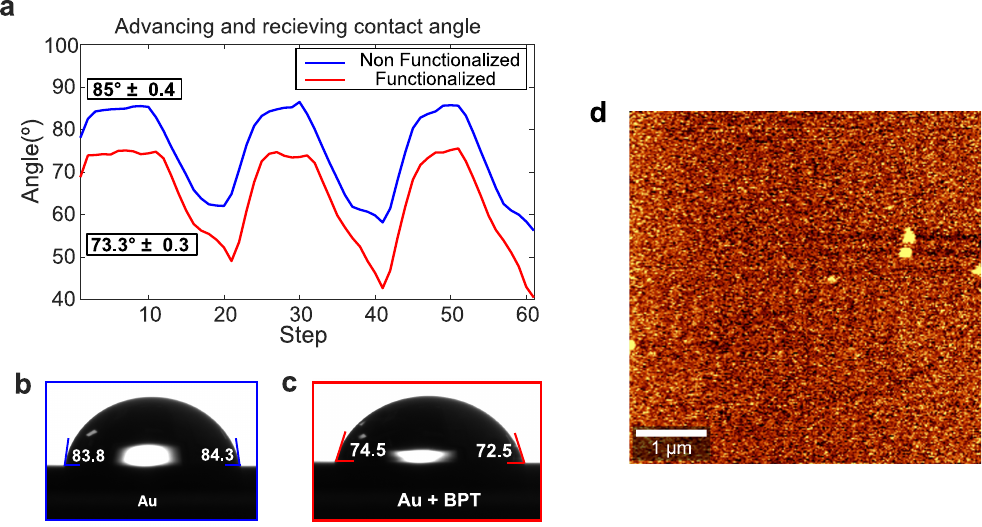}
 \caption{\textbf{Experimental contact angle measurements.} \textbf{a} Advancing and receding comparation between an Au bare surface (blue line) and a BPT-functionalized Au surface (Red line).\textbf{b} Optical image of the dropped surface onto the bare Au sample and the measured angle. \textbf{c} Optical image of the dropped surface onto the BPT-functionalized sample and the measured angle. \textbf{d} AFM image of the deposited SAM.}
 \label{fig_ca_supp}
\end{figure}

An Alpha300 RA (Raman-AFM) instrument from WITec was used to perform AFM measurements, being all measurements conducted in the AC mode. Sharp, uncoated silicon probes (K $\sim$ 42 N/m, $f_0$ $\sim$ 320 kHz) were obtained from PPP-NCH (Nanosensors). All AFM images were processed using WSxM software from Nanotec Electrónica S.L \cite{horcas:2007}. Supplementary Figure \ref{fig_ca_supp} shows an AFM topography image of a BPT functionalized surface, revealing a flat and homogeneous appearance (RMS = 0,75).

\section*{S3. Supplementary simulations}\label{sup_s3} 

In our experiments, we are pursuing the excitation of the fundamental resonance (the 1,0) mode) of the nanoparticle on the metal \cite{kongsuwan:2019}. The excitation of the (1,0) mode of the cavity corresponds to the condition where maximum field confinement is located just beneath the NP, forming a very intense hotspot. To ensure that we are exciting the right mode (the (1,0)), we placed three probes in the region between the NP and the metal. In particular, there is a central probe (number 2) and two additional probes spaced a distance of 1/4 NP from the central probe. Figures a, b, e, and f of Supplementary Fig. \ref{supp_simulations} show the EF as a function of the wavelength at the different probes. The central probe is greater than at probes 1 and 3, which leads us to conclude that the (1,0) mode is excited for both TM and TE modes in both NPs, with the enhancement being greater in NP1, as expected. Figures in the main text show the EF at the position of the central probe (number 2). 


In addition, Supplementary Fig. \ref{supp_simulations} c and d show snapshots of the horizontal $E_x$ and vertical $E_y$ electric field components when the TE mode at $\lambda$ = 690 nm is coupled with the slot waveguide, illustrating how the field is directed to the plasmonic cavity. Supplementary Fig. \ref{supp_simulations}c shows the $E_x$ component, showing clearly its propagation first through the dielectric slot and then through the plasmonic slit, where it finally couples to the plasmonic gaps. If we have a look at the $E_y$ component (Supplementary Fig. \ref{supp_simulations}d) for TE excitation, we see that this component vanishes at the exact cut in the dielectric waveguide region, but it appears in the plasmonic region to excite the plasmonic gaps. The same analysis was made for the TM mode at a $\lambda$ = 671 nm (Supplementary Fig. \ref{supp_simulations}g and h). In this case, both components $E_x$ and $E_y$ are advancing along the dielectric slot, and then along the plasmonic slit antenna to finally excite the (1,0) cavity mode.

\begin{figure}[H]
\centering
\includegraphics[width=\textwidth]{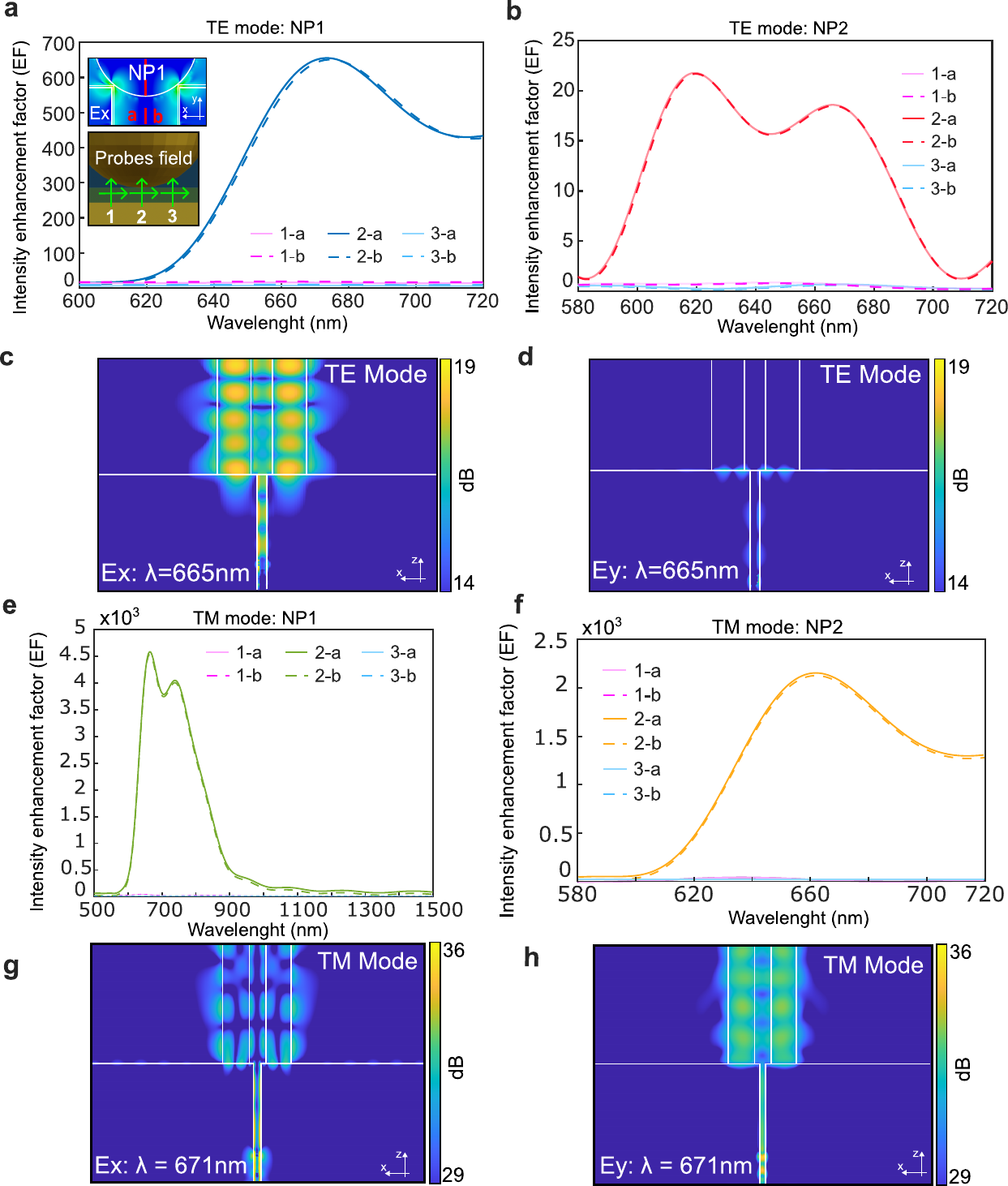}
\caption{\textbf{Numerical results for the evaluation of the excitation of the (1,0) cavity mode and propagation of the electric field components in the visible range for both TE and TM modes}. Enhancement EF at the probes placed in the NP1 \textbf{a} and NP2 \textbf{b} gaps when the structure is excited with the TE mode. The top inset shows how the field is confined in the gap between the NP and the Au patch. The bottom inset shows how the probes are distributed. Top-view snapshot of the propagation of the $E_x$ \textbf{c} and the $E_y$ \textbf{d} component of the electric field for TE excitation at $\lambda$ = 665 nm. Enhancement EF at the probes placed in the NP1 \textbf{e} and NP2 \textbf{f} gaps when the structure is excited with the TM mode. The top inset shows how the field is confined in the gap between the NP and the Au patch. The bottom inset shows how the probes are distributed. Top-view snapshot of the propagation of the $E_x$ \textbf{g} and the $E_y$ \textbf{h} component of the electric field for TM excitation at $\lambda$ = 671 nm.}
\label{supp_simulations}
\end{figure}


\subsection*{Setup}\label{sup_setup} 

The Raman experimental measurements were carried out with the WITec Spectrometer alpha300 RA (Raman-AFM) equipment. Some additional elements were added for sample placement (sample positioner), and a USB microscope (8 LED USB 2.0 Digital Microscopio) for monitoring light coupling into the waveguides. The commercial system consists of an optical line for a laser of $\lambda=633nm$, which includes a light source, polarizers, and mirrors. Then, the Raman signal is collected back by the same objective and taken to a spectrometer through filters and mirrors (see Supplementary Fig. \ref{set_up}.a). The incident light is focused on the sample using an objective for the experimental measurement. The light incidence is vertical; therefore, an additional piece was placed to hold the sample vertically, allowing the waveguides to be oriented in the same direction as the laser propagation. (Supplementary Fig. \ref{set_up}.b)

\begin{figure}[h] \centering
 \includegraphics[width=\textwidth]{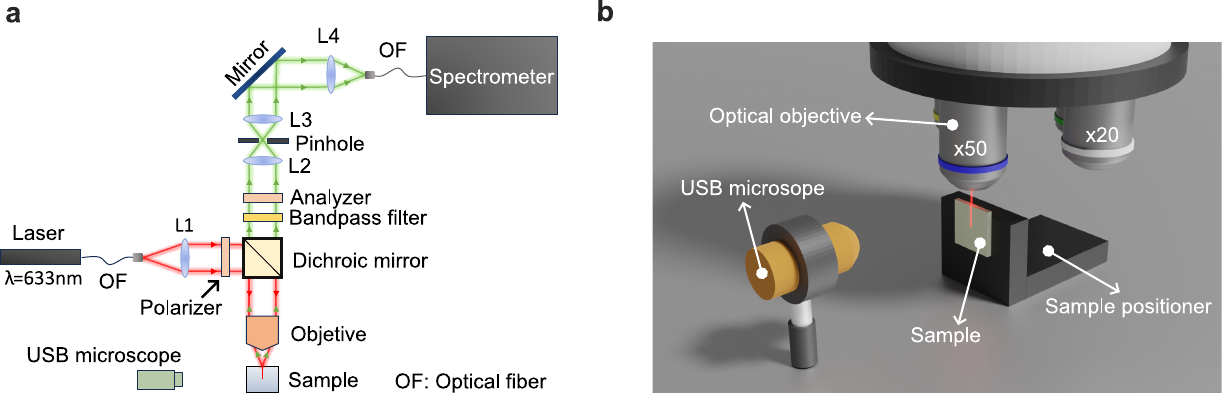}
 \caption{\textbf{Experimental setup configuration and a representation of how the Raman measurements are performed}. \textbf{a} Scheme of the complete optical system where the red lines represent the excitation light path and the green one the collected back Raman signal. \textbf{b} Illustration of the sample positioning and light coupling.}
 \label{set_up}
\end{figure}


\end{appendices}
\bibliography{biblio}
\end{document}